# A Structural Probe of the Doped Holes in Cuprate Superconductors

P. Abbamonte[1,2,¶,♣], L. Venema[1], A. Rusydi[1], G. A. Sawatzky[1,§], G. Logvenov[3], and I. Bozovic[3]

[1] *Materials Science Centre, University of Groningen, 9747 AG Groningen, The Netherlands*

[2] *National Synchrotron Light Source, Brookhaven National Laboratory, Upton, NY, 11973*

[3] *Oxxel GmbH, Bremen, D-28359, Germany*

♣ To whom correspondence should be addressed. Email: peter@bigbro.biophys.cornell.edu

An unresolved issue concerning cuprate superconductors is whether the distribution of carriers in the CuO$_2$ plane is uniform or inhomogeneous. Because the carriers comprise a small fraction of the total charge density and may be rapidly fluctuating, modulations are difficult to detect directly. Here we demonstrate that in anomalous x-ray scattering at the oxygen K edge of the cuprates the contribution of carriers to the scattering amplitude is selectively magnified 82 times. This enhances diffraction from the doped holes by more than $10^3$, permitting direct structural analysis of the superconducting ground state. Scattering from thin films of La$_2$CuO$_{4+\delta}$ ($T_c$ = 39 K) at T=(50 ± 5) K on the reciprocal space intervals (0,0,0.21) → (0,0,1.21) and (0,0,0.6) → (0.3,0,0.6) show a rounding of the carrier density near the substrate suggestive of a depletion zone or similar effect. The structure factor for off-specular scattering was less than $3\times10^{-7}$ $e$, suggesting an absence of in-plane hole ordering in this material.

Soon after the discovery of high temperature superconductivity it was shown that holes doped into an antiferromagnet are unstable to the formation of charged magnetic domain lines or "stripes" with incommensurate spacing (*1,2*). Recent evidence from scanning tunneling microscopy (STM) suggests that the integrated local density of states (LDOS) at the surface of $Bi_2Sr_2CaCu_2O_{8+x}$ is indeed inhomogeneous (*3,4*), and when exposed to an external magnetic field spontaneously forms incommensurate "checkerboard" patterns in the vortex core (*5*). It is an issue central to research on cuprate superconductors to determine if the local carrier density, i.e. the density of doped holes, correlates spatially with these spectroscopy effects. There is therefore a tremendous need for a <u>scattering</u> probe of the doped holes, i.e. which could reveal their distribution in the ground state.

Incommensurate structural features have been observed in the cuprates with both x-ray and neutron diffraction (*6*). However, since the carriers comprise a very small fraction of the total charge density – about 1 part in 500 for optimally doped $La_2CuO_{4+\delta}$ – x-ray scattering is normally dominated by the core electrons. Neutrons couple only to nuclei and magnetic moments and are not sensitive to charge at all. So the relationship between these observations and the distribution of carriers in the ground state, unfortunately, is unclear.

In this report we show that by exploiting resonance effects in the pre-edge part of the oxygen K shell of the cuprates the amplitude for elastic x-ray scattering from doped holes can be selectively magnified by a factor of 82. This enhances diffraction from modulations in the carrier density by more than $10^3$. This technique is similar to MAD crystallography (*7*) or resonant magnetic x-ray scattering (*8*), in which resonance effects provide sensitivity to atomic species or the local magnetic moment, respectively. Like other scattering probes this technique is bulk sensitive (penetration depth ~ 200 nm) and can detect fluctuating phenomena, making it kinematically complementary to STM which detects static, surface effects. The spatial resolution is modest ($\lambda/2$ ~ 1.1 nm), however it is well-suited to mesoscopic phenomena. This is, by many standards, the first direct structural probe of the ground state in the cuprates.

For these measurements we constructed a five-circle diffractometer for scattering soft x-rays. This system mimics the functionality of a single crystal diffractometer but operates in a vacuum of $2\times10^{-10}$ mbar (*9*). The specimens studied were thin films of oxygen-doped $La_2CuO_{4+\delta}$ grown epitaxially on $SrTiO_3$ (*10,11*). Samples with atomic-scale perfection were essential for cleanly demonstrating resonance effects.

The resonant enhancement is a soft x-ray spectroscopy effect which can be understood as follows. The high $T_c$ materials exhibit Cu*L* (933 eV) and O*K* (534 eV) edges, which correspond to removal of an electron from Cu2*p* and O1*s* inner shells, respectively. In x-ray absorption spectra (XAS) of the insulating cuprates the O*K* edge shows a "prepeak" feature (538 eV) which is an intersite O1*s* $\rightarrow$ Cu3*d* transition (*13*) brought about by *p-d* hybridization. The Cu*L* edge comprises two enormous peaks (Figure 1B), which are on-site, dipole-allowed 2*p* $\rightarrow$ 3*d* transitions where the hole has *j* = 1/2 or 3/2 (*13*). Because the valence states are polarized in the $CuO_2$ plane both the oxygen prepeak and the Cu*L* features are observable in XAS only when **E**∥ab (*14*).

When the system is doped, added holes go into O2*p* states (*15*) and a <u>second</u> prepeak, which we will refer to as the "mobile carrier peak" (MCP), appears at 535 eV (Figure 1, red circles). Its oscillator strength builds rapidly with hole concentration as states near the Fermi level are vacated, and also as spectral weight is <u>transferred to 535</u>

eV from the feature at 538 eV (*13*). A transfer of spectral weight with doping is the classic signature of a correlated system (*16*) and it makes the optical response at 535 eV extremely sensitive to the local hole density.

This sensitivity influences diffraction because the amplitude for light scattering from a material is proportional to the dielectric susceptibility, $\chi(\mathbf{k},\omega)$ (*17*). $\chi$ is related to XAS via the absorption coefficient $\mu = 2\,k\,\text{Im}(n)$ where $n = \sqrt{(1+\chi)}$ (SI units) (*18*). We can quantify the size of the resonance at the MCP by calculating $\text{Im}[n(\omega)]$ from $\mu$, Kramers-Krönig transforming to get $\text{Re}[n(\omega)]$, and computing $\chi$. Using the tabulated atomic scattering factors to set the absolute scale, we have determined $\chi$ for optimally doped $La_2CuO_{4+\delta}$ (Fig. 2).

If the carrier density is inhomogeneous the MCP will be modulated in real space, enhancing diffraction of 535 eV x-rays. In carrier-depleted regions the MCP will be nonexistent and $\chi$ will have its pre-resonance value. In carrier-rich regions the peak will be enhanced. To quantify the difference we define a contrast variable, $\Delta \equiv |(\chi_{doped} - \chi_{insulating})/\chi_{insulating}|$, which far from resonance is determined by the electron density to be $\sim 1/500$. According to Figure 2, at the MCP $\Delta$ has the value 0.16. In other words, at 535 eV the scattering amplitude from doped holes is enhanced by $0.16 \times 500 = 82$, increasing the scattered intensity from variations in hole density by $(82)^2$ or more than $10^3$.

The resonance at the Cu$L$ threshold, incidentally, is truly enormous. Defining $\chi_{CuL}$ as the susceptibility at 933 eV and $\chi_{offres}$ as the value 20 eV below the edge (see Fig. 2B), the quantity $|(\chi_{CuL} - \chi_{offres})/\chi_{offres}| \sim 0.78$. In terms of anomalous scattering factors this gives $\Delta f \sim 136$. This feature is not strongly doping dependent so it does not aid our study of the carriers, but it demonstrates how large resonance effects in the soft x-ray regime can be.

The carrier enhancement can be demonstrated in the energy- and angle-dependent reflectivity of a $La_2CuO_{4+\delta}$ film (Fig. 3). In this measurement one might expect interference fringes with angular maxima at $2d \sin(\theta) = n\lambda$. However at $\omega = 533$ eV, labeled "a" in Fig. 1A, fringes are hardly visible (Figure 3A, blue circles). This is because off-resonance the optical constants are determined by the electron density, and the density the $SrTiO_3$ substrate (5.12 g/cm$^3$) and the strained film (< 6 g/cm$^3$) do not differ significantly. So optically the film/substrate ensemble looks like a semi-infinite layer with reflectivity given by the Fresnel formulae. However, if the energy is changed by only 2 eV to the MCP (point "b" in Fig. 1A), there is a 16% boost to the film susceptibility and pronounced interference fringes appear (Fig. 3A, red circles).

The striking thing about these fringes is that they come <u>only from the carriers</u> and are not related to the rest of the electrons in the film. If there were carrier ordering along the (001) direction it would be reflected as a modulation of the envelope of these fringes. Ordering parallel to the $CuO_2$ planes would result in scattering in an off-specular geometry, i.e. in which the momentum transfer vector **q** has a component in the in-plane direction.

The enormity of the Cu$L$ effects are also demonstrable in this measurement. At the top of the Cu$L_{3/2}$ peak the interference fringes, which are barely visible off resonance, approach $10^6$ photons/sec (Fig. 3B, green line). Reflectivity over the full angular range for several x-ray energies is shown in Fig. 3B.

Because the prepeak and Cu*L* edge features are polarized, and we have P polarized light, one expects the interference fringes to grow toward higher scattering angles where **E** lies along the CuO$_2$ planes. The Cu*L* fringes at 933 eV do exactly that. The fringes at the MCP, unexpectedly, <u>attenuate</u> at higher angles. This can be understood in terms of a smoothing of the carrier density at the substrate. In general what makes a layer exhibit interference fringes is two localized discontinuities (the front and rear faces) that generate waves which beat against one another as the film is rotated. The fringes will damp if one or both of these discontinuities is smeared over a depth larger than $\lambda/2$. Several things could potentially do this in our case: (*i*) a carrier depletion region due to the contact potential difference between the film and substrate, (*ii*) oxygen interdiffusion at the substrate, or (*iii*) structural reconstruction at the film-substrate interface induced by the polar nature of a (001) surface.

Without specifying the cause, the effect of smoothing can be demonstrated in the Born approximation in which the scattering amplitude can be expressed in terms of one-dimensional integrals of the form

$$\chi(q) = \int_{-\infty}^{\infty} dx \chi(x) e^{iqx} \qquad (1)$$

where $\chi(x)$ is the spatially-varying susceptibility and $q = 4\pi/\lambda \sin(\theta)$ is the momentum transfer. For an isotropic film of thickness $L$ with susceptibility $\chi_F$ on a substrate with susceptibility $\chi_S$ this integral has the value

$$\chi(q) = \frac{\chi_F}{iq}(1 + e^{-iqL}) - \frac{\chi_S}{iq + \eta} \qquad (2)$$

where a convergence factor $e^{-\eta x}$ attenuates the substrate at infinite distances. If the film is anisotropic the susceptibility is a tensor

$$\chi(q) = \begin{pmatrix} \chi_{ab}(q) & 0 \\ 0 & \chi_c(q) \end{pmatrix} \quad (3)$$

in which case the scattering matrix element, incorporating polarization effects, has the form

$$M = \varepsilon_1 \cdot \chi(q) \cdot \varepsilon_2 \quad (4)$$

Where $\varepsilon_1$ and $\varepsilon_2$ are the incident and scattered polarizations. Taking $\chi_S = 1$, $\chi_F^c = 1.03$, and $\chi_F^{ab} = 1.16$ at the prepeak and $\chi_F^{ab} = 1.78$ at the Cu$L_{3/2}$ peak, we plot $|M|^2$ in Fig. 4 for both a sharp film profile and one that has been smoothed over 20 Å at the substrate. For a sharp layer the fringes at the carrier prepeak increase at high angles as expected. When the interface is smoothed they fade away. So the absence of fringe enhancement at high angles is suggestive of a depletion zone or similar effect at the interface.

In an effort to corroborate the measurements of Ref. (*3*) we scanned the off-specular reciprocal space interval $(0,0,0.6) \to (0.3,0,0.6)$. The sample temperature was $(50 \pm 5)$ K, which is above $T_c$. The elastic scattering signal was below our fluorescence background of 1300 Hz. Using the specular measurements as a reference and scaling, this indicates a structure factor for doped holes on this interval of less than $3 \times 10^{-7}$ electrons, suggesting an absence of in-plane carrier modulation in this material. Whether inhomogeneity is induced below $T_c$ or exists in other cuprates remains to be determined.

¶ Present address: Physics Dept., Cornell University, Ithaca, NY, 14853-2501

§ Present address: Dept. of Physics and Astronomy, University of British Columbia, Vancouver, BC, 46T 1Z1, Canada

18. Here $k = 2\pi/\lambda$ where $\lambda$ is the x-ray wavelength. $n(\omega)$ is the refractive index.

19. We gratefully acknowledge the assistance of S. Hulbert, G. Nintzel, C.-C. Kao, and D. L. Feng, and discussions with L. H. Tjeng, I. Elfimov, J. M. Tranquada, and J. C. Davis. This work was supported by an NSF IRFAP postdoctoral fellowship (INT0002581) and by the NWO (Dutch Science Foundation) via the Spinoza program. Work at the NSLS is supported by the D.O.E. under Contract No. DE-AC02-98CH109886

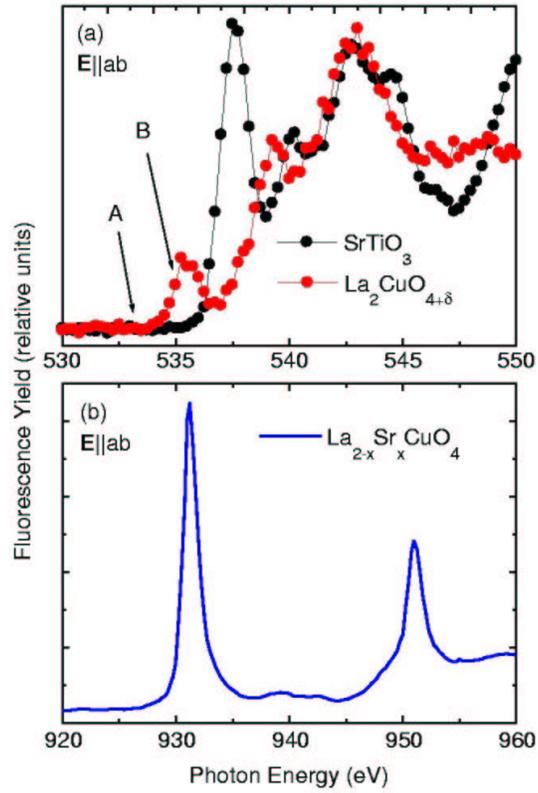

FIG 1 X-ray absorption spectra of optimally doped $LaCuO_{4+\delta}$ and $SrTiO_3$, taken in Fluorescence Yield mode (A) in the vicinity of the Oxygen $K$ edge, showing the mobile carrier peak set off from other features, and (B) at the Cu$L$ edge. All spectra are for **E**∥ab.

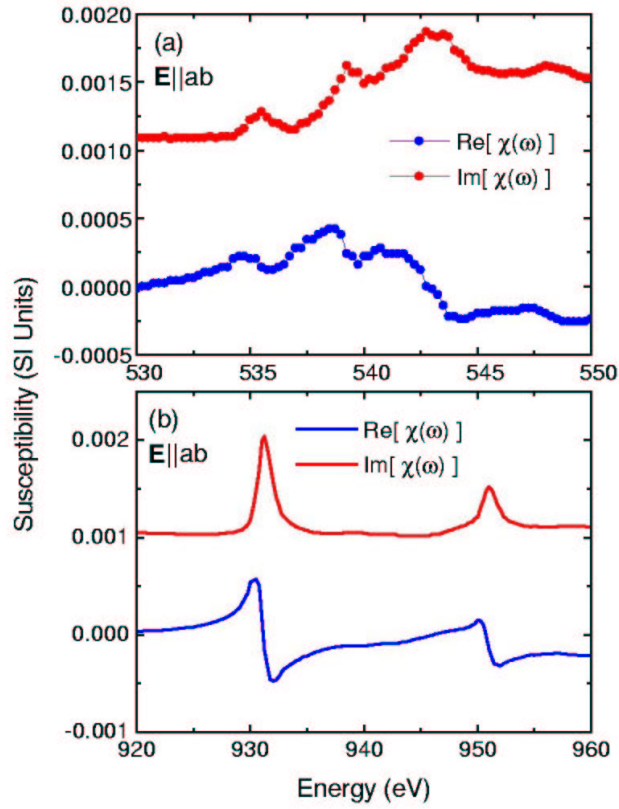

FIG 2 ab-plane susceptibility of optimally doped $La_2CuO_{4+\delta}$, in SI units, calculated by the procedure described, (A) in the vicinity of the Oxygen K edge and (B) the $CuL_{2,3}$ edge.

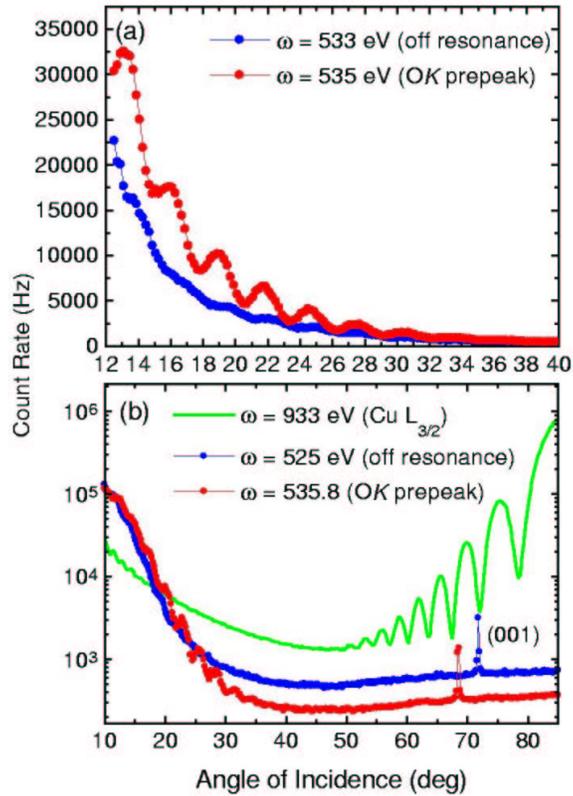

FIG 3 Reflectivity of a 23.2 nm $La_2CuO_{4+\delta}$ film ($T_c = 39$ K) (A) just below and on the mobile carrier peak at 535 eV, showing the resonant enhancement, and (B) over the full angular range for three different energies. The 525 eV and 535.8 eV scans correspond to the reciprocal space interval $(0,0,0.21) \rightarrow (0,0,1.21)$, while 933 eV scan corresponds to $(0,0,0.37) \rightarrow (0,0,2.15)$. The feature at $70°$ is the (001) Bragg reflection.

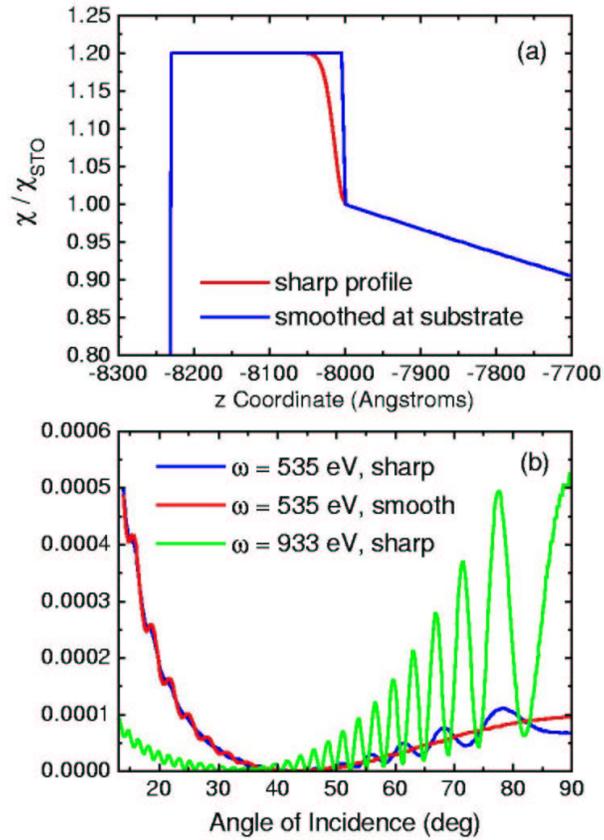

FIG 4 (A) Susceptibility profiles for a perfect film on a substrate (blue line) and one that has been smoothed at the interface (red line). The sloping nature of the substrate (x > -8000 Å) is a numerical implementation of the factor $e^{-\eta x}$. (B) Interference fringes resulting from the two profiles (relative units) showing damping at high scattering angles for the smoothed layer. Formatting is for comparison to Fig. 3B.